\documentclass[12pt]{iopart}
\usepackage{setspace,graphicx,dsfont,verbatim,paralist,indentfirst}
\usepackage[usenames, dvipsnames]{color}
\usepackage{cite}
\usepackage{url}

\begin{document}

\title{Effects of transients in LIGO suspensions on searches for gravitational waves}

\author{M. Walker,$^{1}$
T.~D.~Abbott,$^{1}$ 
S.~M.~Aston,$^{2}$  
G. Gonz{\'a}lez,$^{1}$
D.~M.~Macleod,$^{1}$
J. McIver,$^{3}$
\noindent
B.~P.~Abbott,$^{3}$  
R.~Abbott,$^{3}$  
C.~Adams,$^{2}$  
R.~X.~Adhikari,$^{3}$
% ...
\medskip\\
\centerline{(LSC Instrument Authors	)}}

\address{
$^{1}$Louisiana State University, Baton Rouge, LA 70803, USA 
 
$^{2}$LIGO Livingston Observatory, Livingston, LA 70754, USA 

$^{3}$LIGO, California Institute of Technology, Pasadena, CA 91125, USA 
 
% ...
}

\ead{mwalk49@lsu.edu}
\vspace{10pt}
\begin{indented}
\item[]
\end{indented}

\begin{abstract}

This paper presents an analysis of the transient behavior of the Advanced LIGO suspensions used to seismically isolate the optics. We have characterized the transients in the longitudinal motion of the quadruple suspensions during Advanced LIGO's first observing run. Propagation of transients between stages is consistent with modelled transfer functions, such that transient motion originating at the top of the suspension chain is significantly reduced in amplitude at the test mass. We find that there are transients seen by the longitudinal motion monitors of quadruple suspensions, but they are not significantly correlated with transient motion above the noise floor in the gravitational wave strain data, and therefore do not present a dominant source of background noise in the searches for transient gravitational wave signals. 

\end{abstract}

\section{Introduction}
\begin{figure}
\centering
\includegraphics[width=\linewidth]{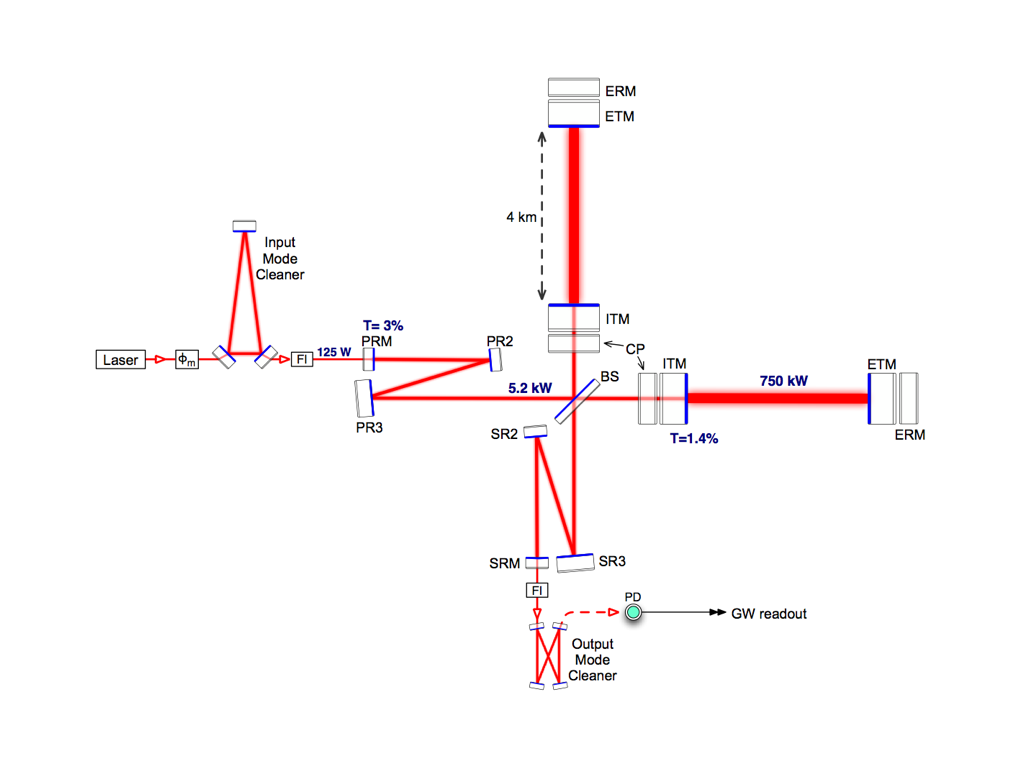}
\caption{Advanced LIGO optical configuration. \cite{aLIGO} The mirrors at the input and end of both arms (labeled ETM and ITM) are suspended from quadruple-stage pendulums, in addition to active seismic isolation systems.}
\label{aligo}
\end{figure}

The Laser Interferometer Gravitational-wave Observatory (LIGO) was designed to detect gravitational waves from astrophysical sources. \cite{LIGO, aLIGO} After six science runs over the course of several years, the detectors underwent major upgrades starting in 2010. In September 2015, the newly upgraded Advanced LIGO detectors began taking data for their first observational period (\textit{O1}). Figure \ref{aligo} shows the basic optical configuration of Advanced LIGO. With a sensitivity more than three times better than that of the previous generation, \cite{DenPaper} the detectors had the astrophysical reach to make the first direct observation of a gravitational wave signal, GW150914, from the merger of two black holes, \cite{GW150914detectionpaper} and later a second unambiguous detection, GW151226. \cite{GW151226} These observations have opened a new field of gravitational-wave astronomy, which will continue to grow brighter with further improvements to the detector network. 

The main low frequency noise source for LIGO is seismic activity. The LIGO detectors are affected by earthquakes from around the world, windy weather that shakes the buildings housing the interferometer instrumentation, microseismic vibrations from ocean waves crashing on the shores of the Pacific Ocean, Atlantic Ocean, and the Gulf of Mexico, and local anthropogenic activity. \cite{EfflerPEM, GW150914detcharpaper} The study of the effects of seismic activity was especially important for improving the data quality of transient gravitational wave searches in the initial LIGO era.\cite{seisveto,s6detchar} One of the key improvements from initial to Advanced LIGO is the implementation of a much more sophisticated seismic isolation system, which includes stages of active and passive isolation for all of the cavity optics.  

It is important to check that this entirely new system provides the very high isolation expected, and that it does not introduce any new types of transient noise that could add to the noise background for searches of short duration gravitational waves, such as black hole mergers and supernovae. Here we present an investigation of the transient motion of the Livingston suspension systems as measured by local sensors on each suspension, specifically looking at the displacement of the quadruple stage pendulums in the longitudinal degree of freedom, which is the direction of the optical path used to sense spacetime strain induced by passing gravitational waves. 

\section{Advanced LIGO Suspensions}

\begin{figure}
\centering
\begin{minipage}{.45\textwidth}
\centering
\includegraphics[width=\linewidth]{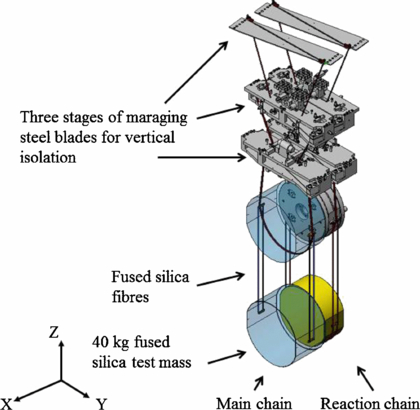}
\end{minipage}
\begin{minipage}{.45\textwidth}
\centering
\includegraphics[width=\linewidth]{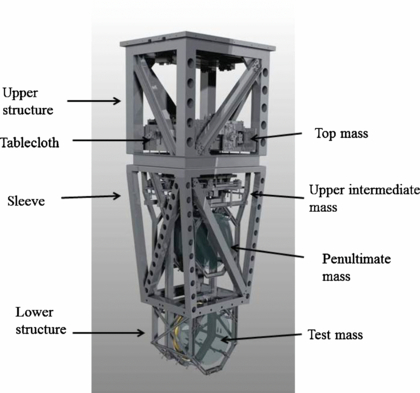}
\end{minipage}
\caption{Quadruple suspensions design. The left image shows the suspension systems with the blades, fibers, and reaction chain. On the right the whole structure is shown, with the four masses labeled. \cite{SuspensionsDesign}}
\label{quadschematic}
\end{figure}

In Advanced LIGO, all optical cavities use optics suspended from multi-stage pendulums, in order to benefit from the lowpassing of seismic motion. The input and end mirrors (the optics whose motion most directly contributes to the gravitational-wave readout signal) are all hung from quadruple-stage suspensions, \cite{SuspensionsDesign} with each stage providing additional isolation at frequencies above the suspension resonances, which range from 0.4 Hz to 14 Hz. The quadruple pendulum is suspended at the top from maraging steel blade springs, with two further sets of springs incorporated into the top two masses, thus providing three stages of enhanced vertical isolation. The two lower masses of the quad are cylindrical silica masses connected by fused silica fibers to reduce thermal noise. Another similar quadruple suspension is hung next to the test mass suspension, so the actuation on lower stages can be done from a similarly isolated reaction chain. Figure \ref{quadschematic} gives an overview of the design of the Advanced LIGO quadruple suspensions.

The local displacement of each stage of the suspensions is measured using Optical Sensor and ElectroMagnetic actuators, or OSEMs, which are electromagnetic sensors and actuators used for damping the suspensions' resonances and controlling the mirrors to keep cavities aligned and locked. \cite{SuspensionsDesign, sensors}

The OSEMs can sense suspension motion at low frequencies where the displacements are relatively large. At frequencies above 5 Hz the suspension motion has typically fallen below the sensitivity level of the OSEMs such that the resulting spectra are dominated by electronics noise. Multiple OSEMs on each stage allow the calculation of the mass's motion in each degree of freedom using a linear combination of the sensors' signals. The sensed displacement of the top stage is used in a feedback loop to actuate on that stage of the suspension in order to damp the mechanical resonances of the suspension. The sensors at lower stages are only used as witnesses of the optics' displacement, for the purposes of diagnosing problems in the suspensions. The actuators on lower stages use interferometer and cavity signals to keep various degrees of the interferometer precisely on resonance.

\section{Motion transients in suspensions}

It is important to understand the origins of non-astrophysical noise transients in the gravitational wave data in order to eliminate false positives from the gravitational wave searches. Each subsystem of the detector itself is therefore investigated in great detail to fully study all potential noise sources. \cite{GW150914detcharpaper}

In this article, we characterize transients in the displacement of the suspensions' stages as measured by the OSEMs, the propagation of transients between different stages, and their effects on the gravitational wave strain data. Specifically, transients in the longitudinal degree of freedom were studied in the top three stages of the quadruple suspensions.

Motion transients seen by the local displacement sensors have a few potential sources. For example, they could be caused by motion that is intrinsic to the suspension systems themselves, from the crackling in the suspension wires or the steel blades. \cite{cracklingnoise, cracklingnoise2}  Transients could also come from excess seismic motion by propagating through each stage of active seismic isolation and then down through the suspension stages. Above the suspension's resonance f$_0$, the seismic transients should decrease in amplitude by a factor of (f/f$_0)^{2}$ at each stage and be less likely to appear above the sensor noise at lower stages. Therefore any seismic transients that affect multiple stages should appear mostly at low frequencies. Another source of transients seen in the local sensors is the actuation on the suspensions from the feedback loops used to control the interferometer.

\subsection{Suspension behavior in Advanced LIGO's first observing run}

\begin{figure}
\centering
\includegraphics[width=\linewidth]{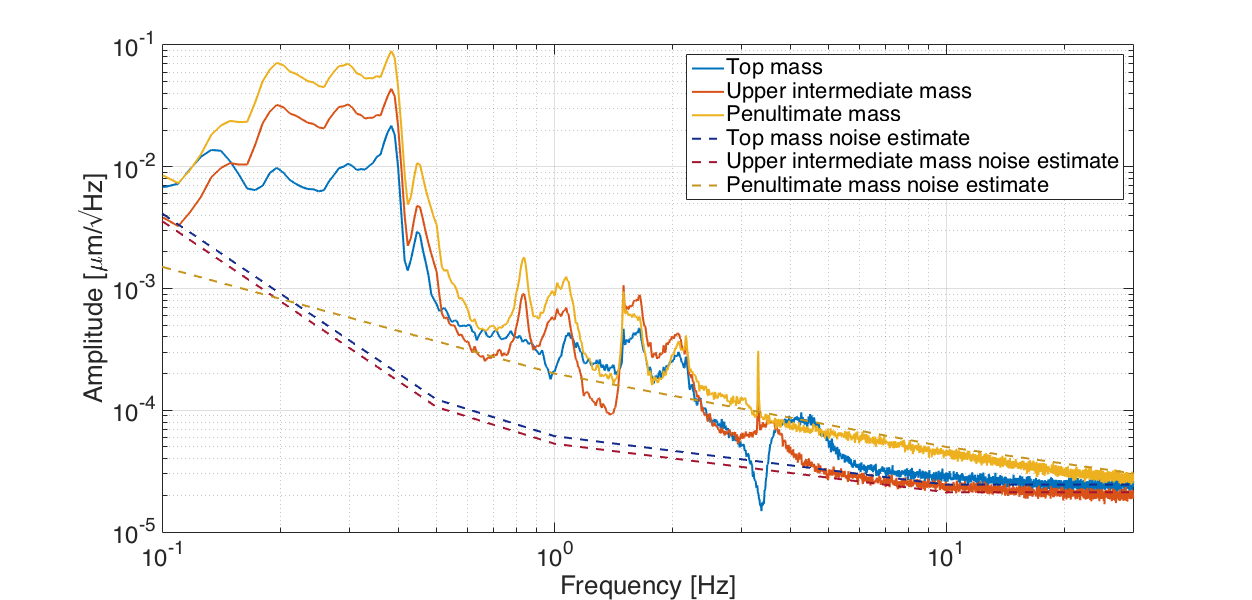}
\caption{Typical amplitude spectral density (ASD) for the ITMY longitudinal motion monitors from O1. Many of the low frequency features of the ASD correspond to the pendulum resonances of the suspension. The flat portion of the ASD above 5 Hz shows where the electronics noise of the OSEM dominates the spectrum. The noise in the penultimate stage at higher frequencies is slightly higher because it has a different kind of OSEM. \cite{sensors}}
\label{ITMY_ASD_AmpvsFreq}
\end{figure}

The typical spectrum of the suspension motion monitors is characterized by several peaks near the low frequency pendulum resonances between 0.4 to 5 Hz, and the flat noise above 5 Hz due to the sensors' electronics noise. The main resonances in the longitudinal degree of freedom are modeled for the quadruple suspensions to be at 0.435 Hz, 0.997 Hz, 2.006 Hz, and 3.416, but coupling from other degrees of freedom and the active seismic isolation system creates additional peaks in the spectrum. Figure \ref{ITMY_ASD_AmpvsFreq} shows a typical spectrum of the Y-arm input quadruple suspension (Input Test Mass Y, or \textit{ITMY}) motion in O1 along with estimated sensor noise levels. 

\begin{figure}
\centering
\begin{minipage}{0.48\textwidth}
\centering
\includegraphics[width=\linewidth]{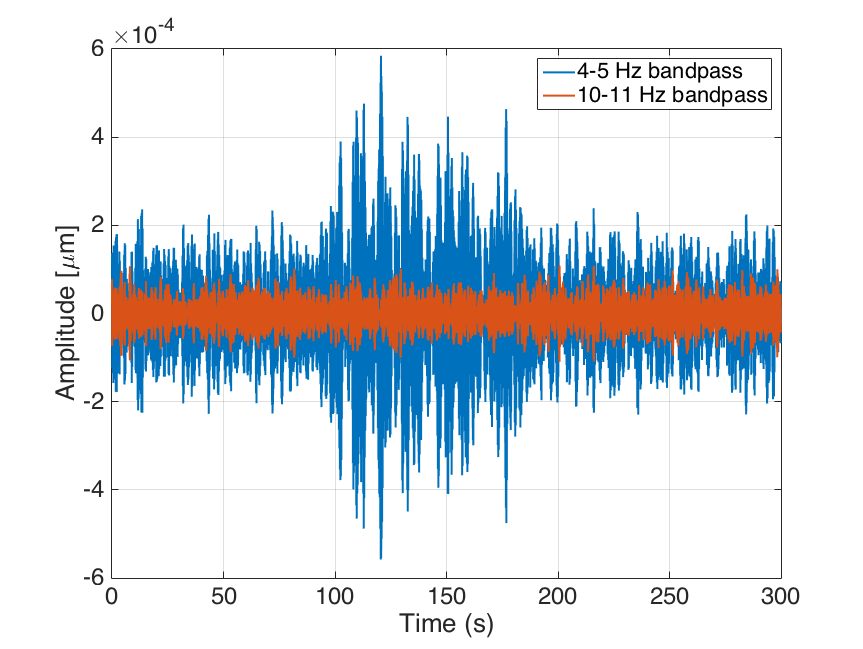}
\end{minipage}
\begin{minipage}{0.48\textwidth}
\includegraphics[width=\linewidth]{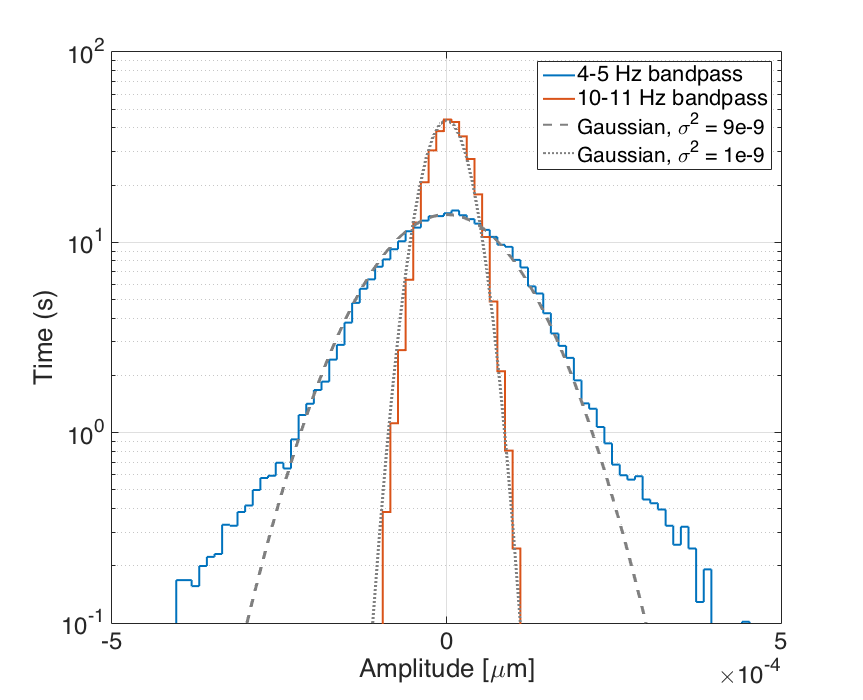}
\end{minipage}
\caption{The panel on the left shows a five minute time series from the top stage of ITMY, with bandpass filters applied between 4 to 5 Hz and 10 to 11 Hz. The distribution of time series amplitude over the same time is shown on the right, with dashed lines to indicate a Gaussian distribution. While the Gaussian-distributed stationary sensor noise dominates the higher frequency band shown, the time series from 4 to 5 Hz exhibits large excursions from the average noise level. }
\label{BandpassedTimeSeries_4to5_10to11}
\end{figure}

Ideally, the noise would be stationary and the average spectrum would statistically characterize the noise level, but in actuality there are non-stationary disturbances at different frequencies. To demonstrate this, Figure \ref{BandpassedTimeSeries_4to5_10to11} shows the time series of several minutes of data from the top stage of one suspension with two different bandpass filters applied to select for 4-5 Hz and 10-11 Hz. Above 10 Hz, the sensor noise dominates the signal and the resulting time series is Gaussian distributed, but the lower frequency shows large non-Gaussian transients.

\begin{figure}
\centering
\begin{minipage}{\textwidth}
\centering
\includegraphics[width=\linewidth]{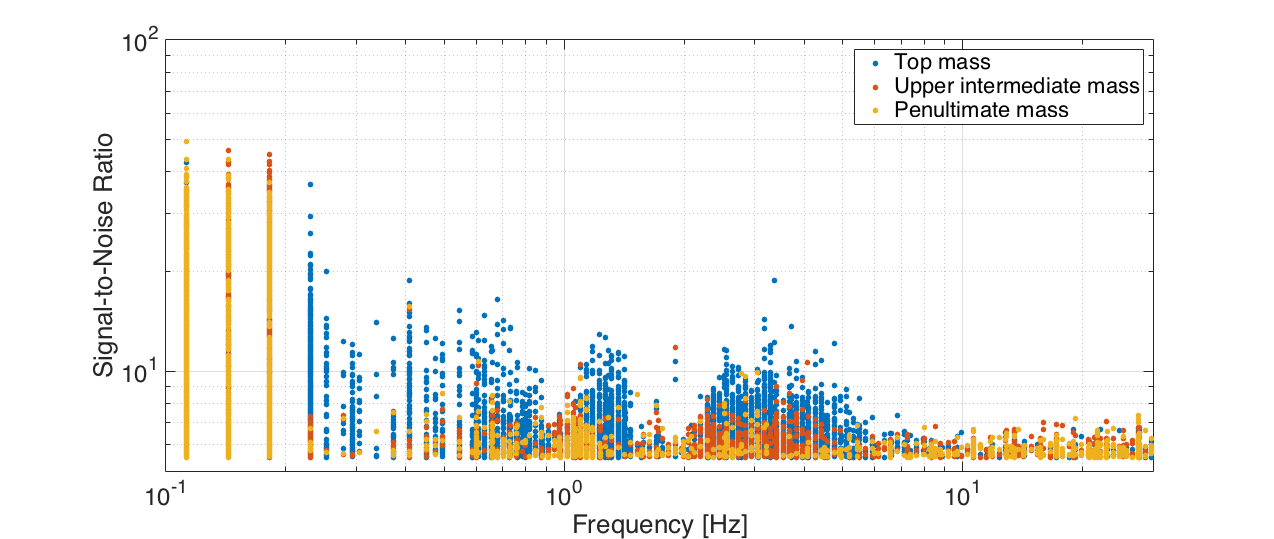}
\end{minipage}

\vspace{20pt}

\begin{minipage}{\textwidth}
\centering
\includegraphics[width = \linewidth]{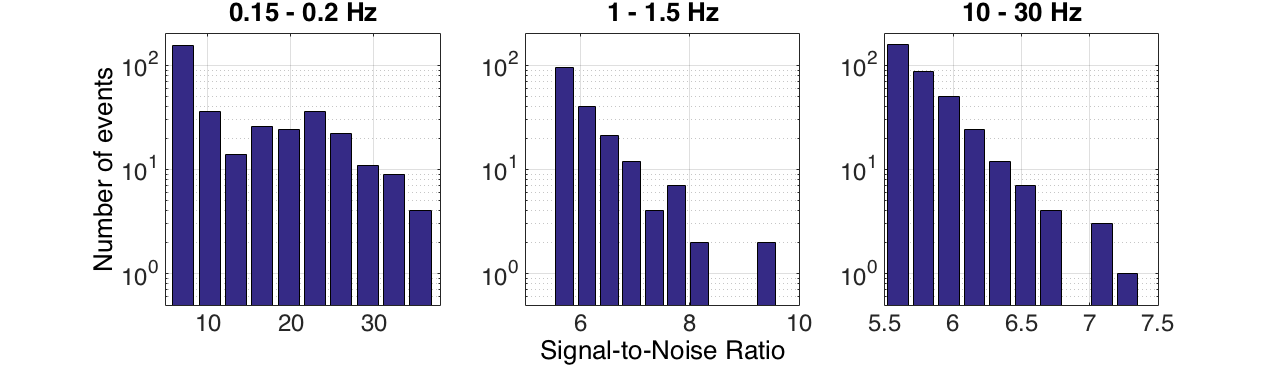}
\end{minipage}
\caption{Above, the Signal-to-Noise Ratio (SNR) and central frequencies of Omicron triggers of ITMY suspension during one week of O1. Below are histograms showing the SNR distributions of the penultimate stage triggers at three selected frequency ranges (note the different SNR scales). While the distribution of higher frequency triggers fall off much like Gaussian noise, the lower frequency ranges contain more outliers.}
\label{OmicronAmpvsFreq_Nov1to8}
\end{figure}

Rather than visually inspecting time series, the Omicron algorithm is used to find transients in the data, producing \textit{triggers} that indicate the time, frequency, amplitude, and signal-to-noise ratio of the transient noise. \cite{OmicronDocument, OmicronArticle} Figure \ref{OmicronAmpvsFreq_Nov1to8} shows distribution in frequency and signal-to-noise ratio of Omicron triggers for the longitudinal degree of freedom, using data from the Y-arm input suspension over one week of the observing run. While stationary noise would produce a background of low SNR triggers across all frequencies, the actual data from the suspension monitors shows a varying structure in different frequency bands. This suggests the presence of non-stationary noise sources. 

\subsection{Motion transient propagation}

\begin{figure}
\centering
\includegraphics[width=\linewidth]{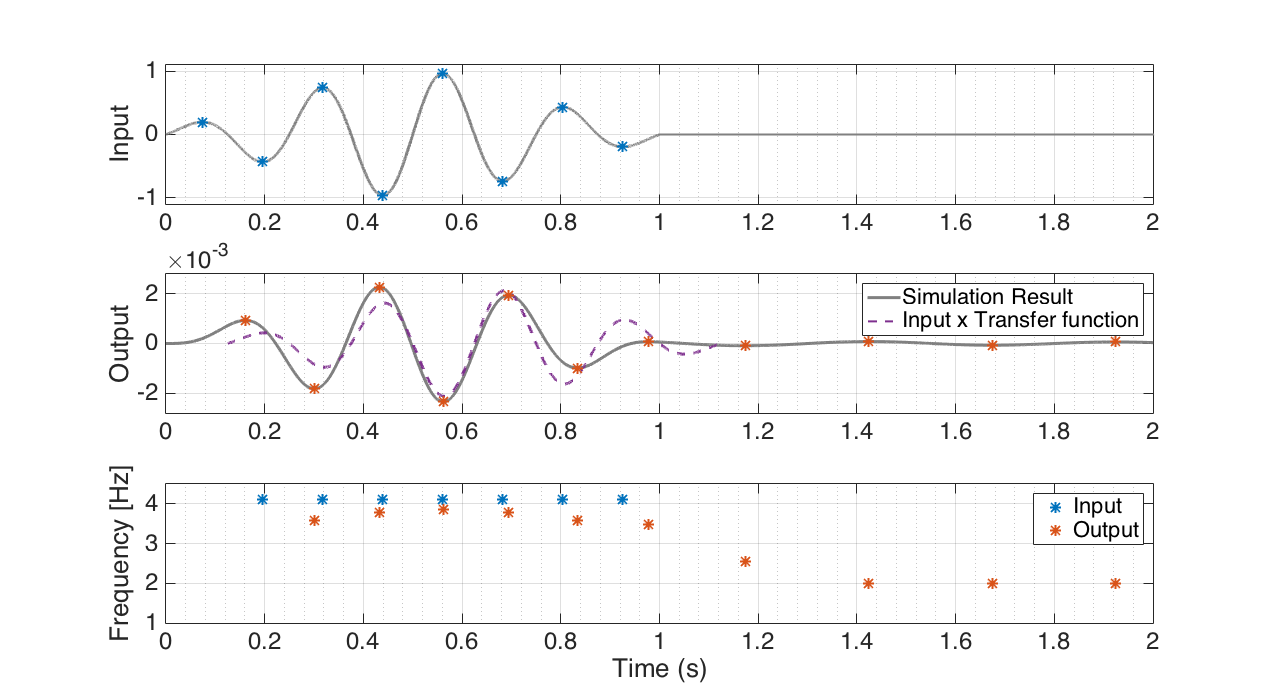}
\caption{Modelled transient response of a simple pendulum to a sine-Gaussian injection. Simulink was used to model a simple pendulum with a resonance at 2 Hz. A one-second 4 Hz sine-Gaussian signal (top panel) was used as the input to show the response of the system (middle panel) compared with the input signal multiplied by the transfer function of the system at 4 Hz. Local maxima and minima of the time series can be used to calculate the frequency for each half-cycle (bottom panel). This simulation shows that even for this simple model, the transient response of the system deviates from the steady state frequency response at 4 Hz.}
\label{SimpleModelResponse}
\end{figure}

To characterize the effects of short duration disturbances in the upper stages of the suspensions on the motion lower in the suspension chain, we need more than just the frequency domain models usually used to characterize the suspensions, due to the influence of the impulse response of the system. Simulink was used to model the response of a simple pendulum to a sine-Gaussian input signal. It is important to note that the input signal is not a pure sine wave at a single frequency, but rather a sine-Gaussian characterized by a peak frequency while also containing broader frequency content. Therefore, the pendulum response more strongly attenuates the higher frequencies of the input signal, and the peak frequency of the resulting motion is a mixture of the driving and the resonance frequencies, as demonstrated in Figure \ref{SimpleModelResponse}.

To examine the propagation of transients in Advanced LIGO suspensions and compare with expected behavior, sine-Gaussian waveforms were physically injected in the Y-end quadruple suspension (End Test Mass Y, or \textit{ETMY}) in the longitudinal direction using the top mass actuators, with central frequencies ranging from 2 to 10 Hz. The Omicron algorithm was used to characterize the resulting transients caused in the top stage as well as in lower stages.

\begin{figure}
\centering
\includegraphics[width=\linewidth]{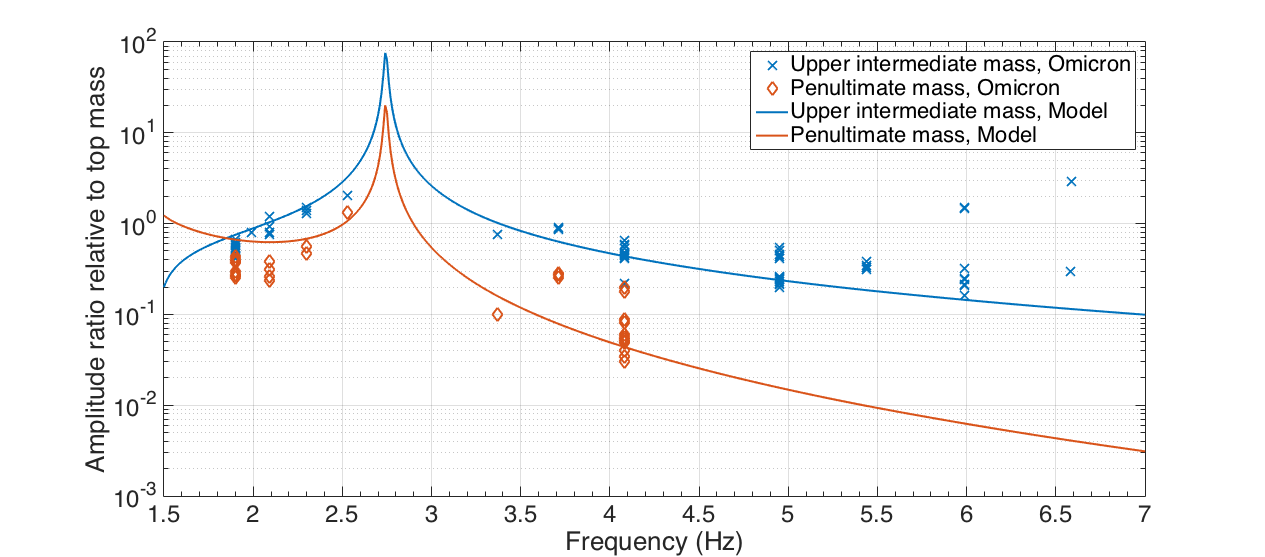}
\caption{Ratio of amplitudes of injection Omicron triggers at lower stages to the top stage, plotted against the peak frequency estimated by Omicron for the top stage motion, compared with the modeled transfer function. At lower frequencies the propagation of the transients is close to the model, but above a few Hz, the motion at the lower stages is smaller than the sensor noise, and the amplitude ratio is not as close to the model. There are fewer Omicron triggers at the penultimate stage, since the motion at that stage is at a lower amplitude and is not great enough at higher frequencies to be seen above the sensor noise.}
\label{AmpRatioInjections}
\end{figure}

Figure \ref{AmpRatioInjections} shows the ratio of the Omicron trigger amplitudes of the lower suspension stages to the top stage for different frequencies, using the Omicron frequency estimate of the top stage trigger. The solid lines show the frequency response of the suspensions as predicted by the quadruple suspension models. One reason for apparent discrepancies from the model is variation of the transient motion frequency between stages, as well as the fact that the motion at each stage is not characterized by only a single frequency. To analyze this effect, time series of the injections were examined individually to characterize the frequencies and amplitudes of the signals at each stage, similar to the process used in analysis of the Simulink model shown in Figure \ref{SimpleModelResponse}. 

\begin{figure}
\includegraphics[width=\linewidth]{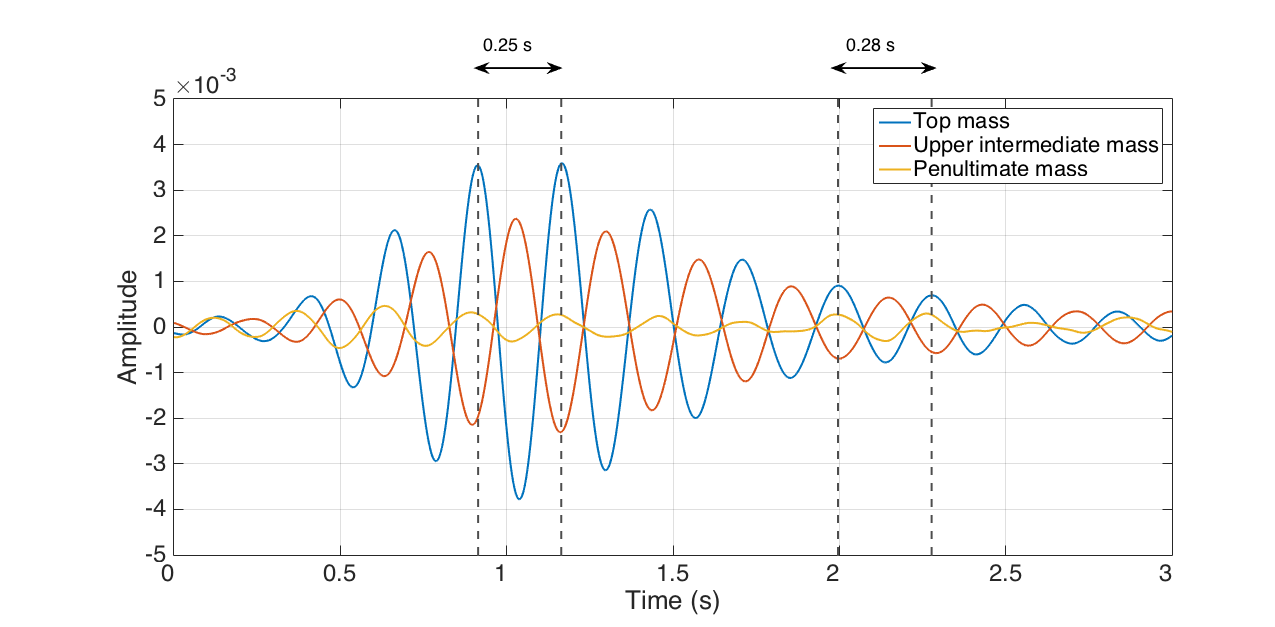}
\caption{Time series from one injection at 4.1 Hz, after application of a bandpass filter with a window of 1 Hz around the injection frequency. Similar to the simple pendulum model analysis, the frequency shifts throughout the time series. The period of the cycles in the top stage lengthens slightly, from 0.25 seconds (4.0 Hz) at the peak of the transient to 0.28 seconds (3.6 Hz) a few cycles later.}
\label{injtimeseries}
\end{figure}

Using a bandpass filter with a 1 Hz window around the injection frequency and finding the local maxima and minima of the resulting time series, the peak frequency of the induced transient motion was estimated with each cycle. Similar to the simulation performed in Simulink, the suspension's response is not exactly at the peak sine-Gaussian frequency, and when the injection is finished the suspension's motion begins to ring down with a frequency approaching the nearest resonance. As the motion propagates downwards, the pendulum filter response attenuates the signal more in the frequency range farther from the resonance, resulting in a slight frequency shift towards the resonance at the lower stage. Figure \ref{injtimeseries} shows the period increasing in the bandpassed time series from one of the injections. 

\begin{figure}
\centering
\begin{minipage}{\textwidth}
  \centering
  \includegraphics[width=\linewidth]{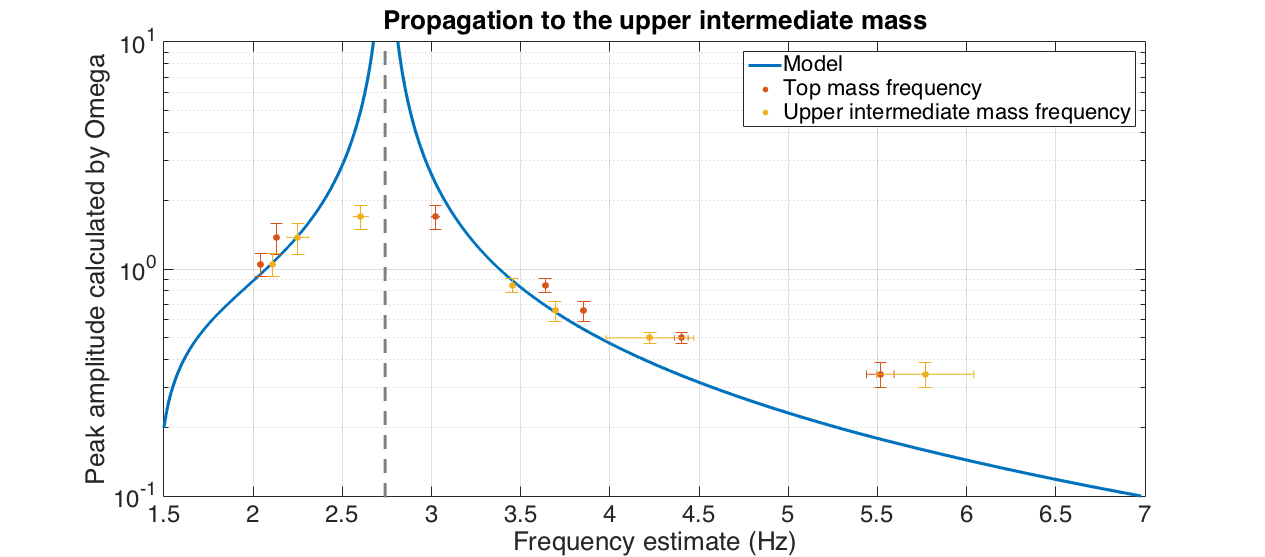}
\end{minipage}
\begin{minipage}{\textwidth}
  \centering
  \includegraphics[width=\linewidth]{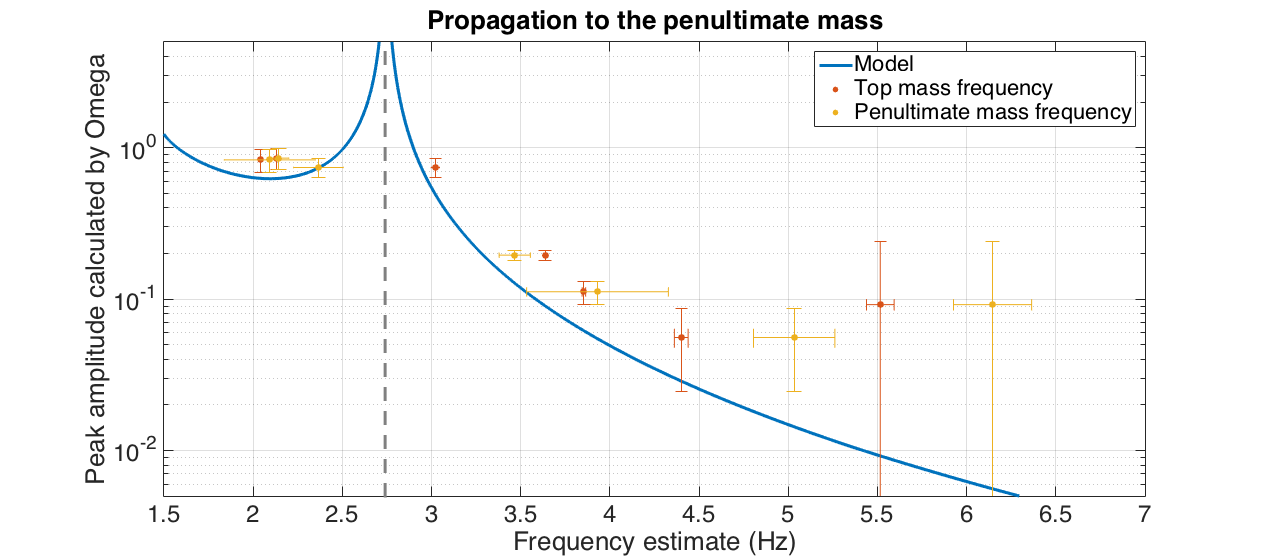}
\end{minipage}
\caption{Amplitude ratios as calculated by the Omega algorithm, and frequency estimated using the maxima and minima of bandpassed time series. Errors are greater as sensor noise becomes dominant at higher frequencies. Red and yellow points represent the same amplitude ratios between stages, but red points show the frequency estimate at the top stage while yellow points show the frequency estimate at the lower stages. Error bars shown are the standard deviation of the measurement among the various injections of the same frequency, weighted by the amplitude of the injection at the top stage.}
\label{InjectionAmpRatiovsFreq}
\end{figure}

The bandpass filter reduces the noise so that the time series cycles can be clearly determined. The frequency of the resulting motion at each stage was then estimated by taking the mean frequency, weighting each cycle by the amplitude of its maximum or minimum. The amplitude of motion was calculated using Omega, a multi-resolution technique for studying transients related to Omicron. \cite{OmicronDocument}\cite{OmicronArticle} Using the weighted average frequency and the amplitudes calculated by Omega, ratio of motion transient amplitudes between suspension stages for each frequency can be better compared to the suspension model. Figure \ref{InjectionAmpRatiovsFreq} displays this comparison for the propagation of the motion from the top stage to the second and third stages. Since the transient amplitude is much smaller at higher frequencies for each successive stage, the higher frequency injection measurements are farther from the model, due to the sensor noise at the lower stages. In both lower stages we see a shift in the frequency away from the frequency at the top stage, generally closer to the nearest suspension resonance, a pitch mode at 2.7 Hz.

Having understood the propagation of short transients in the suspension stages, we turn now to studying the effect of the actual suspension transients on the LIGO gravitational wave strain data during the first Advanced LIGO observing run. 

\subsubsection{Correlations with gravitational wave strain data}

\begin{figure}
\centering
\begin{minipage}{0.48\textwidth}
  \centering
  \includegraphics[width=\linewidth]{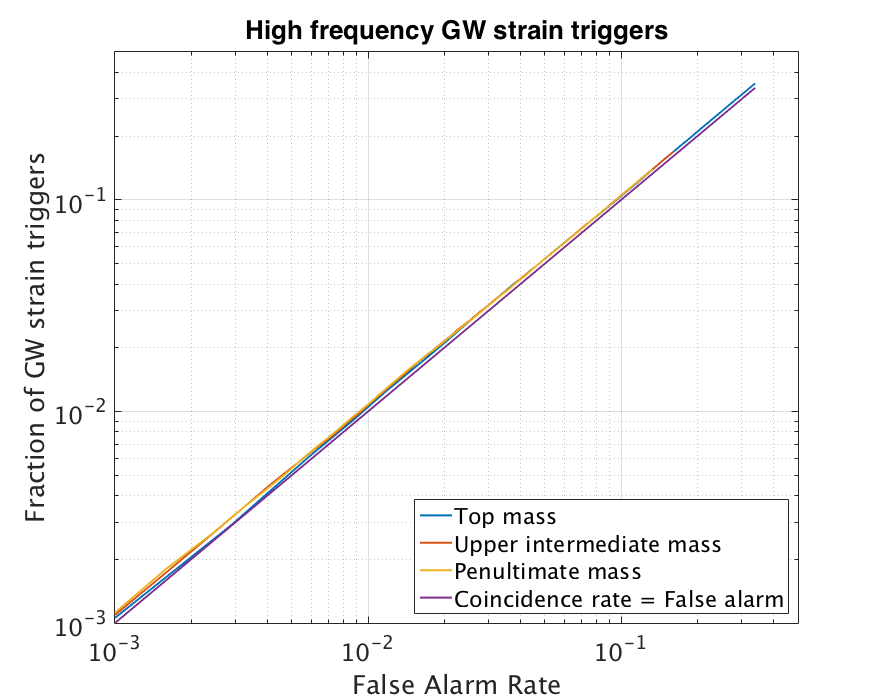}
\end{minipage}
\begin{minipage}{0.49\textwidth}
  \centering
  \includegraphics[width=\linewidth]{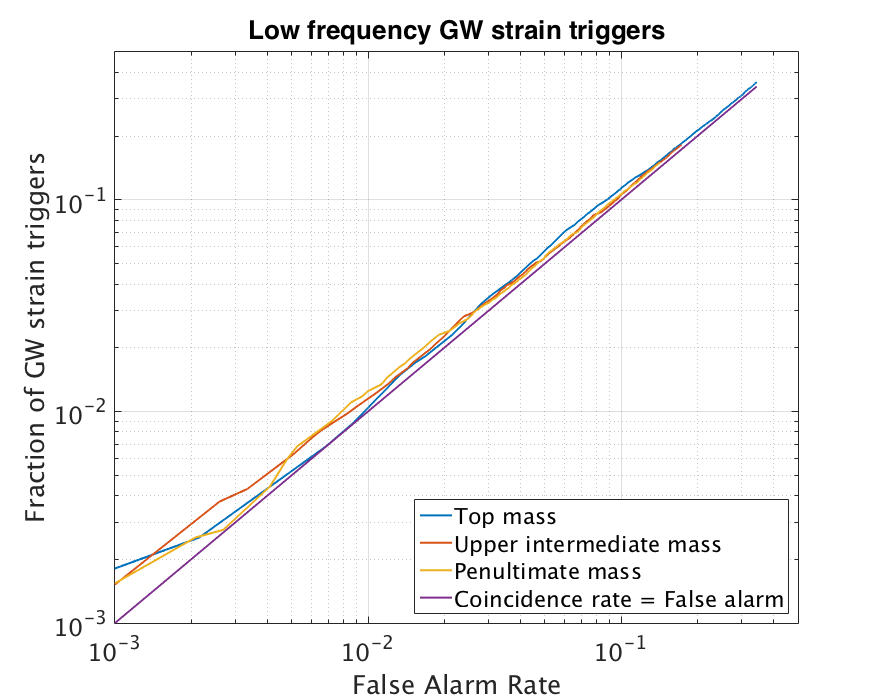}
\end{minipage}
\caption{Receiver Operator Characteristic (ROC) curves showing the correlation between noise transients in the GW strain and ITMY suspension data from November 1 to 8, 2015. The lefthand plot shows the correlation with higher frequency GW strain triggers (above 60 Hz), while the plot on the right shows the correlation with GW strain triggers below 60 Hz. The y axis shows the fraction of triggers coincident between the two sets of data for varying time windows. The x axis represents the number of coincidences that would appear by chance, estimated by repeating the analysis at each time window with different time shifts between the two data sets. For both sets of GW strain triggers, the coincidence rate is approximately equal to the false alarm rate, whereas a significant correlation would have a much greater efficiency than false alarm rate.}
\label{DARMITMYROC}
\end{figure}

Taking the data from the first week of November (the same week shown in Figure \ref{OmicronAmpvsFreq_Nov1to8}), Omicron was used to identify transients in the gravitational wave (GW) strain data in the same frequency range as used to produce the suspension motion triggers (0.1 to 60 Hz), as well as at higher frequencies to check for any nonlinear coupling. Figure \ref{DARMITMYROC} shows the correlations between transients in the ITMY longitudinal displacement data and the gravitational wave strain data in both frequency ranges.  The figures shown are Receiver Operator Characteristics (ROC) curves, which show the time coincidence rate between the two sets of triggers, with various coincidence windows from 0.1 to 10 seconds. This rate is compared with the number of time coincidences that would occur by chance (false alarm rate), using a number of time shifts between the two data sets. In both cases, the small number of coincidences between the sets of data are consistent with the number that would be expected by random chance. The observed transients in the ITMY suspension motion monitors did not show any significant correlation with GW strain noise transients, at any frequency. 

\begin{figure}
\centering
\includegraphics[width=\linewidth]{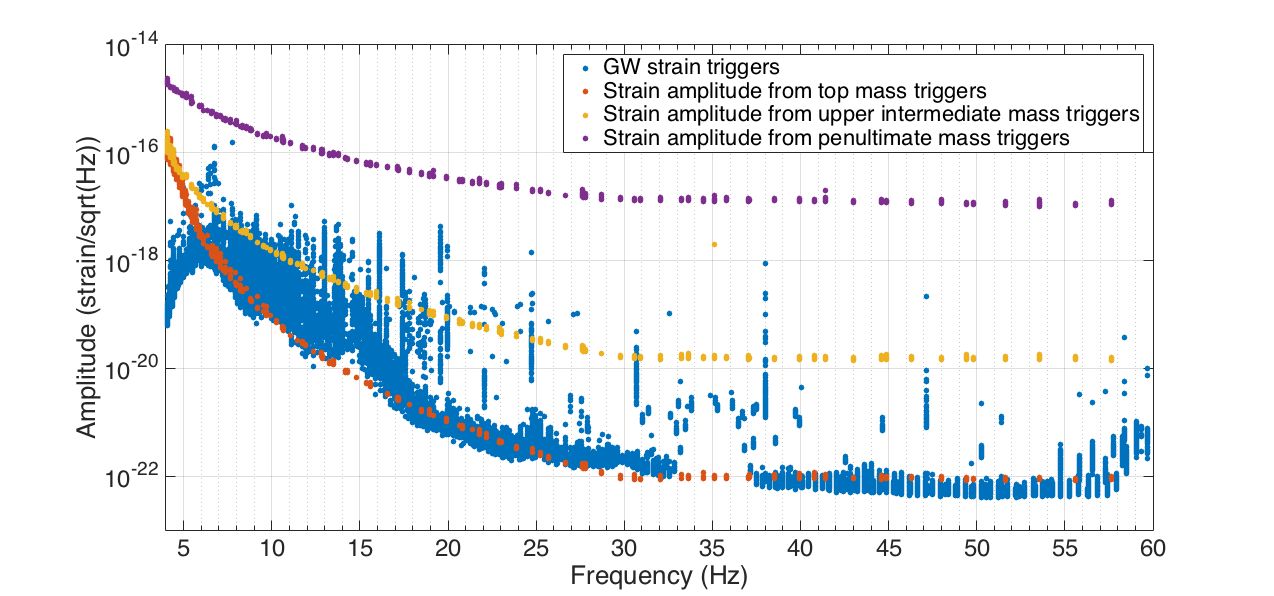}
\caption{Livingston ITMY triggers from a week in the first observing run, multiplied by the transfer function to the lowest stage and divided by the arm length to convert the displacement into equivalent strain amplitude. The strain calibration of the gravitational wave data is only accurate above 10 Hz, at frequencies where the OSEM signals are dominated by sensor noise. Therefore, this calculation can only give us the upper limit of transient motion from each stage that could appear in the GW strain data without also appearing in the local displacement sensor. Where there are GW strain noise transients above one of these levels, we can rule out an origin in a particular stage of the suspension chain. A large number of the GW strain triggers are above the noise level from the top stage, eliminating the origin of the noise at the top of the suspension chain. However, only the very loudest GW strain triggers are above the level of the second stage, and no GW strain triggers are higher than the level of the third stage.}
\label{DarmPredictions}
\end{figure}

We can now place upper limits on the level of noise that would be caused in GW strain from the observed transients in suspension monitors. The amplitude of the Omicron triggers from each of the upper stages of ITMY is multiplied by the suspension transfer function to estimate the amplitude of noise transients that would be caused in the test mass by a physical displacement of that amplitude. Figure \ref{DarmPredictions} shows the resulting projections in equivalent GW strain amplitude, alongside the GW strain triggers from the same time. The sensor noise at the lower stages is much higher than the expected amplitude of motion at those stages, so the upper limit of motion at the lowest stage is above most of the GW strain triggers. The noise level predicted by the top stage triggers, however, is below most of the GW strain triggers up to 37 Hz, so if noise originating in that stage caused high amplitude transients in the GW data, it would be expected to also be seen by the top stage sensors. 

Since the top stage triggers are not statistically correlated with any of the GW strain triggers, we can conclude that transient noise originating at the top stage of the suspension is not a significant contribution to the transient noise in the interferometer. We cannot, however, rule out the possibility of GW strain noise transients caused by motion originating in the lower stages of the suspension, since there are a significant portion of GW strain triggers that fall below the level of transient noise caused by the local sensor noise.

\section{Conclusions}

Using short duration hardware injections in the top stage of the suspension, we have studied the propagation of transient motion down the suspension chain. The difference of transient amplitudes at different stages is consistent with the models, although slight variations in frequency must be taken into account. The frequency of the transients shifts because the injected waveform is not a pure sine wave but a sine-Gaussian, and after the short duration injection, the suspension motion oscillates with a decreasing amplitude and frequency that shifts toward the closest mechanical resonance frequency. Transients at different stages of the suspension therefore show slightly different frequencies from the same initial sine-Gaussian injection. 

Statistical comparisons of the times of transients in the OSEMs and in the GW strain data during O1 show that transients seen by the local displacement sensors of the suspensions are not a significant source of background transient noise in the interferometer. However, this does not rule out transient suspension motion that is below the local sensor noise as a possible source of background noise. Using the suspension models to propagate the sensor noise into the motion at the test mass, upper limits can be placed on the level of noise that could be caused in the GW strain data from transients in suspension motion at each stage. Most GW strain triggers are above the sensor noise level of the top stage of the suspension, but below the noise level of the third stage. Transient noise that originates in the lower stage of the suspension could therefore be a cause of noise in the GW data while not being loud enough to appear above the local sensor noise. \cite{aLIGO}

\ack
LSU authors acknowledge the support of the United States National Science Foundation (NSF) with grants PHY-1505779, 1205882, and 1104371.

The authors gratefully acknowledge the support of the NSF for the construction and operation of the
LIGO Laboratory and Advanced LIGO as well as the Science and Technology Facilities Council (STFC) of the
United Kingdom, the Max-Planck-Society (MPS), and the State of
Niedersachsen/Germany for support of the construction of Advanced LIGO 
and construction and operation of the GEO600 detector. Additional support for Advanced LIGO was provided by the Australian Research Council.
The authors also gratefully acknowledge the support of LSC related research by these 
agencies as well as by 
the Council of Scientific and Industrial Research of India, 
Department of Science and Technology, India,
Science \& Engineering Research Board (SERB), India,
Ministry of Human Resource Development, India,
the Istituto Nazionale di Fisica Nucleare of Italy, 
the Spanish Ministerio de Econom\'ia y Competitividad,
the  Vicepresid\`encia i Conselleria d'Innovaci\'o, Recerca i Turisme and the Conselleria d'Educaci\'o i Universitat del Govern de les Illes Balears,
the European Union,
the Royal Society, 
the Scottish Funding Council, 
the Scottish Universities Physics Alliance, 
the Hungarian Scientific Research Fund (OTKA),
the National Research Foundation of Korea,
Industry Canada and the Province of Ontario through the Ministry of Economic Development and Innovation, 
the Natural Science and Engineering Research Council Canada,
Canadian Institute for Advanced Research,
the Brazilian Ministry of Science, Technology, and Innovation,
International Center for Theoretical Physics South American Institute for Fundamental Research (ICTP-SAIFR), 
Russian Foundation for Basic Research,
the Leverhulme Trust, 
the Research Corporation, 
Ministry of Science and Technology (MOST), Taiwan
and
the Kavli Foundation.
The authors gratefully acknowledge the support of the NSF, STFC, MPS and the
State of Niedersachsen/Germany for provision of computational resources. 
\section{References}
\bibliographystyle{unsrt}
\bibliography{paper}

\end{document}